\documentclass[10pt]{article}
\usepackage[left=1.5cm,top=2cm,right=1.5cm,bottom=2.5cm,nohead]{geometry}
\input amssym.def
\input amssym.tex

\usepackage{graphicx}

\def\nn{\nonumber}

\def\ben{\begin{equation}}
\def\een{\end{equation}}
\def\half{{1 \over 2}}
\def\bea{\begin{eqnarray}}
\def\eea{\end{eqnarray}}
\def\babla{{\mbox{\boldmath $ \nabla $ }}}

\def\br{{\bf r}}\def\bx{{\bf x}}
\def\by{{\bf y}}

\def\bn{{\bf n}}

\def \p{\partial} \def \half{\frac{1}{2}}

\begin{document}

\hfuzz=100pt
\title{Geometrical and physical models of abrasion}
\author{G. Domokos \\
{\em Department of Mechanics, Materials, and Structures }\\
{\em Budapest University of Technology and Economics}\\
{\em M\"uegyetem rkp.3 }\\
{\em Budapest 1111, Hungary} \\
{\em and }  \\
G.~ W.~ Gibbons  \\
{\em D.~A.~M.~T.~P.}\\
{\em  Cambridge University}\\
{\em Wilberforce Road, Cambridge CB3 0WA, U.K.}\\}

\maketitle

\begin{abstract}

We extend the geometrical theory presented in \cite{DG} for collisional and frictional particle abrasion
to include an independent physical equation for the evolution of mass and volume.
We introduce volume weight functions as multipliers of the geometric equations and
use these mutipliers to enforce physical volume evolution in the unified equations.
The latter predict, in accordance with Sternberg's Law, exponential
decay for volume evolution. We describe both the PDE versions, which are generalisations
of Bloore's equations and their heuristic ODE approximations,
called the box equations. The latter are suitable for tracking the 
collective abrasion of large particle populations. The mutual
abrasion of identical particles,  called the self-dual flows, play a key
role in explaining geological scenarios. We give stability
criteria for the self-dual flows in terms of the parameters of the physical volume evolution models
and show that under reasonable assumptions
these criteria can be met by physical systems.
We also study a natural generalisation, the Unidirectional Bloore
equation,  covering the case of unidirectional abrasion.
We have previously shown that his equation admits travelling
front solutions with circular profiles.  
More generally, in three dimensions,  they are so-called linear or special
Weingarten surfaces.

\end{abstract}

\pagebreak

\tableofcontents

\pagebreak 

\section{Introduction}

In our earlier paper \cite{DG} we investigated Bloore's collisional partial differential equation (PDE) \cite{Bloore}
describing the evolution of particle shapes under isotropic collisions:
\ben \label{Bloore}
-v= a(1+2bH +cK)
\een
where $v$ is the evolution speed in the direction of the inward normal, $a={\rm constant}$ with the dimension of speed, $H= \half (k_1+k_2)$ is the mean curvature
and $K=\kappa_1\kappa_2$ is the Gauss curvature and $b$ and $c$
are constants with the dimensions of length and   ${\rm length}^2$
respectively. In \cite{DG} we approximated (\ref{Bloore}) by a set
of ordinary differential equations called the box equations under the assumption that all shapes are ellipsoidal
and remain so for all times, i.e. it is sufficient to track the evolution of the orthogonal bounding boxes. 
The box model was successfully tested against laboratory experiments and recently against a
detailed field study along the Williams river, Australia \cite{Williams}.

In the current paper we extend and generalise our previous work. The original Bloore equation (\ref{Bloore}) and its box approximations
correctly describe  the evolution of geometrical shapes, 
however, these are purely geometrical equations and thus
unable to predict the correct time evolution for mass and volume. 
One important sign of this shortcoming is
that the model (\ref{Bloore}) predicts finite lifetimes for all 
particles whereas field observations in fluvial environments
indicate an exponential decay as formulated by Sternberg's empirical formula, also called Sternberg's Law \cite{Sternberg}. This indicates that
volume evolution has to be derived from physical equations independent of the Bloore model.

Although physically incorrect, the Bloore model (and its box approximations) still predict volume evolution rates depending on the normal
speed $v$ from (\ref{Bloore}) and on the geometry of the surface $\Sigma$:
\ben \label{geom}
\dot V^g(v) =\int _{\Sigma}vdA \,.
\een
where the superscript $g$ refers to the geometrical equations and $\dot()$ denotes differentiation with respect to time.
These rates we call \it the geometrical volume evolution \rm and  
we derive the exact formulae in section \ref{s:volgeom}.
As we can see in (\ref{geom}), $\dot V^g$ is a linear function of the normal speed $v$ in (\ref{Bloore}), i.e. 
\ben \label{linear1}
\dot V^g(\lambda v)=\lambda \dot V^g.
\een
Subsequently, in section \ref{s:volumeweight} in the spirit of Firey's work \cite{Firey} we introduce the \it volume weight functions \rm 
$f(V(t))$ which depend only on time and do not depend on the location on the surface.
These functions enter Bloore's equation instead of the constant $a$ and we also define their analogues in the box equations.
If we have an independent physical model for volume evolution predicting volume diminution rate
$\dot V^p$ (the superscript referring to the independent physical equations) then we can set this  equal to the 
volume diminution predicted by the volume-weighted geometrical equations
\ben
\dot V^g(f(V)v)=\dot V^p
\een
and this condition yields, via the linear property (\ref{linear1}) 
\ben
f(V)=\frac{\dot V^p}{\dot V^g}.
\een
This illustrates that volume weight functions can be used
to suppress the geometrical volume evolution rates entirely in favour of the physical ones. After introducing in section \ref{s:collective}
the basic equations for 
the statistical theory of collective abrasion, 
in section \ref{s:phys} we introduce some models which predict physical volume diminution in accordance with
Sternberg's Law, so combining these models with the original geometrical equations via the volume
weight functions yields shape and size evolution consistent both with the geometrical Bloore theory
as well as Sternberg's  empirical formula for volume diminution.

In addition to introducing the volume weight functions and the physical volume evolution into the geometrical model,
we also generalise the original Bloore model in other ways. In section \ref{s:mutual} we introduce the coupled system of PDEs describing
the mutual abrasion of two particles, as well as the box 
approximations of these equations. All previously
mentioned equations deal with collisional abrasion which, as we pointed out in \cite{DG} is not capable
on its own to adequately describe the collective evolution of pebbles in geological environments.
In section \ref{s:friction}  we introduce the PDE including friction and also its box approximations. 
In section \ref{s:phys} we also provide the
physical volume evolution model for the frictional case. 

Frictional abrasion is particularly significant, because in \cite{DG} we showed that in the box
flows if identical shapes mutually abrade each other (which we call the self-dual flow) then
friction may stabilise nontrivial shapes as global attractors. However, it was not clear whether
these shapes are also attractive in size, i.e. whether the self-dual flows are stable
with respect to perturbations in size. Earlier we pointed out that global transport
resulting in size segregation may stabilise these flows. While that is certainly a valid possibility,
in section \ref{s:stab} we show that a potentially more relevant mechanism is defined by the physical models
of volume diminution. In the models introduced in our current paper we show the exact condition under which
a physical volume diminution model can stabilise the self-dual flows.   

Beyond isotropic particle abrasion we also discuss unidirectional abrasion and in Appendix \ref{s:wein}
we show that under such conditions linear Weingarten surfaces emerge as translationally invariant solutions
of the unidirectional Bloore equation.

The current version of the manuscript is intended to convey both the theoretical PDE models based on Bloore's equation
as well as to provide detailed basis for a computer code simulating collective abrasion based on the box equations.
The latter could serve as a platform to compare these results with field data and laboratory data.
Due to this double goal, readers interested in any one of the above subjects may find some equations which
appear less relevant to their immediate purpose. On the other hand, separation of the two subjects also
raises difficulties and at this stage we decided to keep the material at least temporarily unified.

\section{Collisional abrasion of an individual particle in constant environment}

\subsection{Bloore's Local  Equation}

In \cite{Bloore} Bloore  proposed that the shape of
the bounding surface $\Sigma$ of pebbles made of a homogeneous material 
and  eroded by a gas of small spherical abraders
should be governed by a  local  equation
of the form
\ben
-v= F(\kappa_1,\kappa_2) \label{UrBloore}\,,
\een
where $\kappa_1,\kappa_2  = \frac{1}{R_1}, \frac{1}{R_2}$
and $R_1,R_2$  are the principal radii of curvatures, $v$ is the speed along the inward normal 
at which the  local area element $dA$ is being eroded
and $F(\kappa_1,\kappa_2)$  is some symmetric function of the principal
curvatures $\kappa_1,\kappa_2$. 
The simplest case is perhaps  (\ref{Bloore}), mentioned in the Introduction.
 For spherical abraders of radius $r$, 
Bloore gave a statistical argument that 
\ben
b=r\,,\qquad c=r^2 \,. 
\een
For non-spherical abraders, a more sophisticated treatment using 
Schneider-Weil theory \cite{VarkonyiDomokos} 
leads to 
\ben \label{coeffs}
b= \frac{M}{4 \pi} \,, \qquad c= \frac{A}{4 \pi} \,, 
\een
where
\ben
M= \int _\Sigma H dA \,,\qquad A = \int _\Sigma  dA
\een
are the integrated mean curvature and area respectively.
Thus one expects on purely
 dimensional grounds that  the first term to be important
for pebbles whose linear size is large  compared with the size of the abraders
while for  pebbles whose linear size is comparable   with the size of the abraders 
the second and third terms should be increasingly important.
Evidently, when the size of the pebble is comparable with the size
of the abraders, the {\it single pebble } treatment like Bloore's
breaks down and the evolution of the abraders must also be considered. 

In the mathematics literature the  three terms in (\ref{Bloore}) 
are often treated separately. The first term in (\ref{Bloore})
\ben
-v= a \label{Eikonal}
\een
is called the {\it Eikonal equation} or the {\it parallel map} 
and arises in the study of wave fronts with speed $a$,
satisfying Huygens's principle. Given an initial aspherical
surface the Eikonal flow  tends to make the surface more aspherical
and to develop faces which intersect on edges \cite{DomokosSiposVarkonyi}. 
   
The second term  in  (\ref{Bloore}) 
 \ben
-v= 2abH \label{Mean}
\een
is called the {\it mean curvature flow}
\cite{Brakke} and often arises  in problems where surface tension is important
\cite{Brakke,RCD}. Given an initial aspherical
surface it tends to make the surface more spherical \cite{Huisken}.

The third  term  in  (\ref{Bloore})
\ben
-v= ac K  \label{Gauss}
\een
is called the {\it Gauss  flow} 
and it also tends to make the surface more spherical
\cite{Tso,Andrews,Andrews1}. 
 
For completeness we mention a fourth flow which is sometimes studied
for its special mathematical properties \cite{Chow} 
which we call the {\it Rayleigh} flow  
\ben \label{rayleigh}
-v= {\rm constant} K^{\frac{1}{4}} \,.
\een
The reason for our name is that this flow
 has the property, first noticed by Lord Rayleigh \cite{Rayleigh1,Rayleigh2,Rayleigh3}
that under it, ellipsoids evolve in a self-similar fashion.

\subsection{Level set representation}

If we describe the moving shape $\Sigma$ as the level sets
\ben
t+ \phi(x,y,z) =0\,,
\een
we may transcribe  a Bloore type  equation 
for the moving surface $\Sigma$   
as a PDE for $\phi(x,y,z)$ as follows.
In one time step
\ben
dt + \babla \negthinspace\phi \cdot d\br =0\,. 
\een
where $\br=[x,y,z]^T$ is the position vector defining the surface.
Thus we have
\ben
1+ \babla \negthinspace \phi\cdot \frac{d\br}{dt} =0 \,. 
\een
But the velocity $v$ in the normal direction is 
\ben
v= \frac{d\br}{dt} \cdot\frac{\babla \negthinspace \phi}{|\babla \negthinspace \phi | }  
\een
Thus
\ben
1+  |\babla  \negthinspace\phi | v =0\,.
\een
where $v=v(\kappa_1,\kappa_2)$ and $\kappa_1,\kappa_2$ may be expressed in terms
of $\phi$ (see e.g. \cite{Goldman}). 
In particular
\bea
H &=& \half \babla \negthinspace\cdot  \frac{\babla \negthinspace \phi}{|\babla \negthinspace \phi | } \,\nn\\ 
K&=& \frac{G}{(\phi_x^2 + \phi_y^2 +\phi_z^2 )^2  } 
\nn\, \\
G&=& \phi_x^2 ( \phi_{yy} \phi_{zz}- \phi_{yz}^2 ) +\phi_y^2 
( \phi_{zz} \phi_{xx}- \phi_{zx}^2 ) + \phi_z^2 ( \phi_{xx} \phi_{yy}- 
\phi_{xy}^2 ) \nn \\ &&
+2 \phi_x\phi_y ( \phi_{xz}\phi _{yz} -  \phi_{xy} \phi_{zz} ) 
+2 \phi_y\phi_z ( \phi_{yx}\phi _{zx} -  \phi_{yz} \phi_{xx} )
+2 \phi_z\phi_x ( \phi_{zy}\phi _{xy} -  \phi_{zx} \phi_{yy} )\,
\label{Gausscurvature} 
\eea
and of course we have
\ben \label{principal}
\kappa _{1,2}=H\pm \sqrt{H^2-K}\,.
\een
\subsection{Monge representation}

Following Monge \cite{Monge}, if $\Sigma$ is a single-valued
function in $(x,y)$ then we may represent it
as a graph over a plane 
\ben
z-h(x,y,t)=0  \,.
\een
Since the normal is $ \frac{1}{\sqrt{1+h_x^2+h_y^2 }} (-h_x,-h_y,1)  $ 
we obtain the Bloore equation as a  PDE in $x,y,t$  
  \ben
\frac{\p h}{\p t}  = \frac{1}{\sqrt{1+h_x^2+h_y^2 }} v  \label{Monge}
\,.\een
\noindent The standard expressions for $H$ and $K$ 
may be obtained by substituting  $\phi=z-h(x,y,t)$
in (\ref{Gausscurvature}).
\bea
K&=& \frac{h_{xx} h_{yy} -h_{xy}^2}{(1 + h_x^2 + h_y^2 )^2 } \,.
\nn \\
H&=& \half \frac{(1+h_y^2)h_{xx} + (1+ h^2_x) h_{yy} - 2 h_x h_y h_{xy} } 
{ (1+h_x^2 +h^2_y)^{\frac{3}{2}} }     
 \,,\eea

An interesting application of both sets of formulae
is to the surface
\ben
xyz=c \,,\qquad \Longleftrightarrow \qquad z=\frac{c}{xy}   
\een 
for which
\bea
K & = & \frac{3 c^3} {(x^2y^2 + y^2 z^2 + z^2 x^2 ) ^2  }\\
H & = & - \frac{c(x^2 + y^2 + z^2 )}{(y^2 z^2  +z^2 x^2  + x^2 y^2) ^{3/2}}
\eea
Interestingly, this family of surfaces is   invariant under
the Rayleigh flow (\ref{rayleigh}) since it is a Titzeica surface, 
that is the stutz or support function $\bx \cdot \bn$ 
is constant multiple of  $K^{1/4}$.

\subsection{Relation to the Kardar-Parisi-Zhang equation}

In soft condensed matter physics, interfaces are
often modelled using the  
the  {\it  Kardar-Parisi-Zhang equation}
for  the height function $h=h(x,y)$ 

\ben
{\p h \over \p t}= \nu \nabla ^2 h  +{\lambda \over 2} 
(\nabla h )^2 + \eta (x,y,t) \label{KPZ}
\een 
where $\nabla$ is with respect to the {\it flat}  metric on ${\Bbb E} ^2$ and
$\eta(x,y,t)$ is a Langevin-type  stochastic Gaussian noise term 
\cite{KPZ,Marsilli}.
It was pointed out in \cite{Maritan} that this was not re-parametrisation
invariant and is an approximation to a   stochastic version
of the {\it mean curvature flow}.  
\ben
v= -\nu H + \lambda + \eta(\sigma ^A,t)  
\een
The first term is essentially the functional derivative 
of surface energy, i.e. a 
{\it surface tension term}  and the
second is the functional derivative 
of  a   volume energy i.e. a  {\sl pressure}  term.
In the absence of the stochastic noise, i.e. if $\eta=0$ and if $
\nu, \lambda >0$,  the system should 
relax to a surface of constant mean curvature $H= \frac{\lambda}{\nu}$.
For pebbles $\lambda = a$ and $ \nu = -2ab$ and the pressure is negative.
In the absence of the noise term, the KPZ equation (\ref{KPZ})
may,by means of the substitution
$w= \exp (\frac{\lambda}{2 \nu} h ) $, reduced  to 
the linear diffusion equation for $w$ \cite{Batchelor}. 
\subsection{Box Equations}

The Bloore equations are partial differential equations
and define a flow on the infinite space of shapes.
In \cite{DG} a finite dimensional truncation was introduced 
which leads to a finite number of ordinary differential equations
referred to as the {\it box equations}. The basic idea is to
bound our pebble by  rectangular box of sides $2u_1,2u_2, 2u_3$ 
ordered such that $u_1\le u_2\le u_3$ 
which defines an inscribed  ellipsoid  of semi-axes $u_1,u_2,u_3$.
One then writes down three equations 
\ben
-\dot u_i= F({\kappa_1}_i,{\kappa_2}_i) 
\een
where $i=1,2,3$ and  ${\kappa_1}_i,{\kappa_2}_i$ are now taken to  
be the curvatures
of the inscribed ellipsoid at the ends of the three  principal axes
$(\pm u_1,0, 0)$, $(0,\pm u_2,0)$,$ (0,0,\pm u_3) $. 
Thus (\ref{Bloore}) takes the form
\ben
-\dot u_1 = a \left( 1+ b \left( \frac{u_1}{u_2^2} + \frac{u_1}{u_3 ^2}\right)
+ c \frac{u_1^2 }{u_2^2u_3^2 }   \right) \,, \qquad {\rm etc} \label{Boxeqn}
\een
where ${\rm etc} $ denotes two further equations obtained by cyclic  permutation
of the suffices $1,2,3$.
 
In \cite{DG} it was found convenient to replace the three lengths
$u_1,u_2,u_3$ by two dimensionless 
ratios and a length $y_1= \frac{u_1}{u_3}$, $y_2= 
\frac{u_2}{u_3}$ and $y_3= u_3$, yielding
\bea 
\label{yy_box3D}
\dot y_i & = & aF_i(y_1,y_2,y_3,b,c) =a\left(\frac{F_i^E}{y_3} +2 b \frac{F_i^M}{y_3^2}  + c \frac{F^G_i}{y_3^3}\right)\, \\
\label{y3_box3D}
-\dot y_3 & = & aF_3(y_1,y_2,y_3,b,c)= a\left(1 + \frac{b}{y_3} \frac{y_1^2+y_2^2}{y_1^2y_2^2}+\frac{c}{y_3^2}\frac{1}{y_1^2y_2^2}\right)\,,
\eea
where
\ben \label{components}
F_i^E= y_i-1, \hspace{0.5cm}F_i^M=\frac{1-y_i^2}{2y_i}\hspace{0.5cm}F_i^G=\frac{1-y_i^3}{y_iy_j^2}, \hspace{0.5cm}i,j=1,2; i\not=j \,.
\een
By introducing the vector notation $\mathbf{y}=[y_1,y_2,y_3]^T, \mathbf{F}=[F_1,F_2,F_3]^T$, 
(\ref{yy_box3D})-(\ref{y3_box3D}) can be rewritten as
\ben
\label{i_individual_box3D}
\dot{\mathbf{y}}=a\mathbf{F}(\mathbf{y},b,c)\,,
\een
which is identical to equations (2.2)-(2.6) of \cite{DG}.

A special case of the box equations are the {\it spherical flows}
for which $u_1=u_2=u_3=R$, where $R$  is the radius of the sphere.
The spherical flows obtained from the box equations  in fact
coincide with the  exact  solutions of the  full
 partial differential equations
(\ref{Bloore}) obtained by assuming that $\Sigma$ is a sphere.
 
\section{Collisional abrasion of two, mutually colliding particles} \label{s:mutual}
 
\subsection{Binary Bloore Equations}

In the Bloore equations the abraders are assumed to be constant
in shape and size.  It is, however, simple to write
down a set of evolution equations  for both
the abraders and the abraded pebbles as done  in \cite{DG}
for the simplified case, the box equations.
In that case we introduced semi-box-lengths
$v_1,v_2,v_3$ for the abrading particles,
yielding  two dimensionless ratios and one length
$z_1=\frac{v_1}{v_3}$, $z_2=\frac{v_2}{v_3}$ and $z_3=v_3$.
Retaining  the notation of \cite{DG}   we use the labels $y$ 
and $z$ for abraded and abraded, by utilising (\ref{coeffs}), the obvious
partial differential equations to consider are  
\bea
\label{BinaryBloore1}
-v_y&=&  a\left(1+2\frac{M_z}{4\pi} H_y  +\frac{A_z}{4 \pi}K_y\right) \,\\
\label{BinaryBloore2}
-v_z&=& a\left(1+2\frac{M_y}{4\pi} H_z  +\frac{A_y}{4 \pi}K_z\right).
\eea

\subsection{Binary Box Equations}

In the box approximation the mean curvature and surface area integrals in 
(\ref{coeffs}) are replaced by the corresponding quantities of the orthogonal bounding box
of the the incoming particle (which, for simplicity is now taken as the $\mathbf{z}$ particle):
\ben \label{integ}
M  = 2\pi z_3(z_1+z_2+1) \,, \quad
A  = 8z_3^2(z_1z_2+z_1+z_2)\,.
\een
The same quantities can be expressed for the unit cube as 
$M_1 = 6\pi, \quad  A_1  =  24$,
so in the box equations we have
\ben \label{box_coeffs}
b(\mathbf{z})= \frac{M}{M_1}  =z_3\frac{z_1+z_2+1}{3}= z_3f^b_z\,,\quad
c(\mathbf{z}) = \frac{A}{A_1}=z_3^2\frac{z_1+z_2+z_1z_2}{3}= z_3^2f^c_z\,.
\een
The corresponding binary box equations  can be written as
\bea
\label{y_box3D}
\dot{\mathbf{y}}&=&a\mathbf{F}(\mathbf{y}, b({\mathbf{z}}),c({\mathbf{z}}))= a\mathbf{F}^c(\mathbf{y},\mathbf{z})\\
\label{z_box3D}
\dot{\mathbf{z}}&=&a\mathbf{F}(\mathbf{z}, b({\mathbf{y}}),c({\mathbf{y}}))=a\mathbf{F}^c(\mathbf{z},\mathbf{y})
\eea
where superscript $c$ refers 
to collisional abrasion. Equations (\ref{y_box3D})-(\ref{z_box3D}) are similar
to  equations (2.13)-(2.14) of \cite{DG}). 

\subsection{The self-dual flows}
As written, the equations (\ref{BinaryBloore1})-(\ref{BinaryBloore2}) have a solution for  which
the abraders and abraded have identical forms. This solution
we refer to as the {\it self-dual flow}. For the self-dual flow
the labels $y$ and $z$ are redundant and we are left with the single equation
\ben
-v= a\left(1+2\frac{M}{4\pi} H  +\frac{A}{4 \pi}K\right)
\label{Selfdual} \een
which in the box approximation reads
\ben \label{Selfdual_box}
\dot{\mathbf{y}}=a\mathbf{F}(\mathbf{y}, b({\mathbf{y}}),c({\mathbf{y}}))= a\mathbf{ F}^c(\mathbf{y},\mathbf{y}).
\een
An important question is whether the self dual
flow (\ref{Selfdual}) or its box version (\ref{Selfdual_box}) are stable within the class of Binary Bloore flows
(\ref{BinaryBloore1})-(\ref{BinaryBloore2}) and Binary Box flows (\ref{y_box3D})-(\ref{z_box3D}), respectively.

\subsection{The spherical case}
If both particles are spherical (with
radii $R_y$ and $R_z$, respectively), then both the binary Bloore equations (\ref{BinaryBloore1})-(\ref{BinaryBloore2}) 
and the binary box equations (\ref{y_box3D})-(\ref{z_box3D}) collapse to the same
two coupled first order ordinary differential equations:  
\bea \label{sphere1}
-\dot R_y&=& a\left(1+2 \frac{R_z}{R_y}  +\left(\frac{R_z}{R_y}\right) ^2\right)                         \\
\label{sphere2}
-\dot R_z&=& a\left(1+2 \frac{R_y}{R_z}  +\left(\frac{R_y} {R_z}\right) ^2\right).
\eea

\section{Frictional abrasion of an individual particle: Non-local theory} \label{s:friction}

\subsection{Bloore equations with friction}

In \cite{DG} the effects of mutual friction, both rolling and sliding
were incorporated into the box equations. This can be done
at the level of the equations describing the  the complete
evolution of the pebble   but while the equations
remain first order in time they become rather non-local
in the coordinates $u,v$ used to parametrise
the embedding
\ben
\br= \br(u,v,t) 
\een
of the surface $\Sigma$ into  Euclidean space.

We define  $R(u,v,t)= |\br(u,v,t) - \bar \br(t)|$ to be 
the distance of the point $\br(u,v,t)$  
from the instantaneous centroid  $ \bar \br(t)$ of the pebble.
We also define $R_{\rm max} (t)$ and $R_{\rm min} (t)$ 
as the instantaneous  maximum and minimum of values
of  $R(u,v,t)$ over the surface and we postulate that frictional
abrasion is governed by
\ben \label{friction1}
{\p \br(u,v,t)  \over \p t}=-G(R, R_{min}, R_{max}) \bn(u,v,t)\,,\qquad   G>0 \,. 
\een 
In \cite{DG} several constraints on the general form of
of the function
$G(R, R_{{\rm min} } , R_{{\rm max}})$ were given and also
one example satisfying these constraints was demonstrated, introducing
separate terms for sliding and rolling with independent coefficients $\nu_s, \nu_r$,
respectively and the dimensionless ratios $r_1=R/R_{min}, r_2=R/R_{max}$:
\ben \label{friction10}
G(R, R_{min}, R_{max})=\nu_sf_s(r_1,r_2)+\nu_rf_r(r_1,r_2)=\nu_sr_2r_1^{-n}+\nu_rr_2(1-r_2^n)\,.
\een
According to the arguments discussed in \cite{DG},
for sufficiently high values of $n$, this model appears to capture most essential physical 
features of frictional abrasion. While (\ref{friction10}) is clearly just an example (\cite{DG} describes 
also an alternative equation), however,
it provides a simple basis for a qualitative analysis.

Frictional abrasion can be readily introduced into the Bloore equations.
As before we use the labels $y$ and $z$.
Since friction is an {\sl additional independent }  
mechanism for abrasion it is natural to assume that 

\bea
-v_y&=& a\left(1+2\frac{M_z}{4\pi} H_y  +\frac{A_z}{4 \pi}K_y\right)
+ G(R_y, R_{y\, {\rm min}} , R_{y\, {\rm max}} )
\,,\\
-v_z&=& a\left(1+2\frac{M_y}{4\pi} H_z  +\frac{A_y}{4 \pi}K_z\right)
+ G(R_z, R_{z\, {\rm min}} , R_{z\, {\rm max} } )\,.
\label{FrictionalBloore}\eea

In case of spherical flows (\ref{friction10}) reduces to a single constant $\nu_s$, so we have

\bea \label{sphere1_f}
-\dot R_y&=& a\left(1+2 \frac{R_z}{R_y}  +\left(\frac{R_z}{R_y}\right) ^2\right) +
\nu_s \,,\\
\label{sphere2_f}
-\dot R_z&=& a\left( 1+2 \frac{R_y}{R_z}  +\left(\frac{R_y} {R_z}\right) ^2 \right)+ \nu_s\,.
\eea

\subsection{Box equations with friction}

If we take the $n\to \infty $ limit in the semi-local PDE (\ref{friction10}) we obtain 
for the box variables
\ben \label{f1}
\dot u_1  = - \nu_s y_1  -  \nu_r y_1,\quad \dot u_2  = -\nu_r y_2, \quad \dot u_3 = 0\,,
\een
where $\nu_s, \nu_r$ are the coefficients for sliding and rolling friction, respectively.
Equation (\ref{f1}) is equivalent to
\ben 
\label{yy_box3D_friction}
\dot \mathbf{y}=\mathbf{F}^f(\mathbf{y},\nu_s,\nu_r) =\frac{1}{y_3}(\nu_s\mathbf{F}^S+\nu_r\mathbf{F}^R)\,,
\een
where
\ben \label{f_components}
\mathbf{F}^S=-\left[ y_1,0,0\right]^T \,, \qquad \mathbf{F}^S=-\left[y_1,y_2,0\right]^T. 
\een
We can now simply add collisional and frictional flows 
(\ref{y_box3D})-(\ref{z_box3D}) and (\ref{yy_box3D_friction})
to obtain the collisional-frictional equations  for the two-body problem:
\bea
\label{yf_box3D}
\dot \mathbf{y} & =&  a\mathbf{F}^c(\mathbf{y},\mathbf{z})+\mathbf{F}^f(\mathbf{y},\nu_s,\nu_r) \\
\label{zf_box3D}
\dot \mathbf{z} & = &  a\mathbf{F}^c(\mathbf{z},\mathbf{y})+ \mathbf{F}^f(\mathbf{z},\nu_s,\nu_r) \,.
\eea

\section{Volume evolution in the geometric equations} \label{s:volgeom}

\subsection{Geometric volume evolution in the Bloore equations: spherical case}
The Binary Bloore equations (\ref{BinaryBloore1})-(\ref{BinaryBloore2}) define the mutual evolution of \emph{observable quantities}, such as
maximal width $D$, surface area $A$ and volume $V$. In general, we can not obtain closed formulae for their evolution, however,
the spherical case admits such computations.
In case of spherical particles with radii
$R_y,R_z$ volume evolution can be derived by integrating (\ref{sphere1})-(\ref{sphere2})
on the surface, to obtain
\ben \label{volumesphere}
-\dot V_y =  -\dot V_z = 4 a\pi(R_y+R_z)^2 
\een
which we call the \emph{geometrical volume evolution} for spheres in the binary Bloore equations.

\subsection{Geometric volume evolution in the box equations}

In the box equations we can derive geometric volume evolution for arbitrary shapes. Regardless whether
the abrasion is collisional or frictional,
the volumes $V_y,V_z$ of the two particles can be expressed as
\bea \label{vol1y}
V_y & = & 8y_1y_2y_3^3 \,,\\
\label{vol1z}
V_z & = & 8z_1z_2z_3^3\,.
\eea
By differentiating (\ref{vol1y})-(\ref{vol1z}) with respect to time we get for the geometric volume evolution:
\bea \label{voly}
\dot V^g_y (\mathbf{y,\dot y})& = & \frac{d}{dt}(8y_1y_2y_3^3)=8\left(\dot y_1y_2y_3^3+y_1\dot y_2y_3^3+3y_1y_2y_3^2\dot y_3\right)\,, \\
\label{volz}
\dot V^g_z (\mathbf{z,\dot z})& = & \frac{d}{dt}(8z_1z_2z_3^3)=8\left(\dot z_1z_2z_3^3+z_1\dot z_2z_3^3+3z_1z_2z_3^2\dot z_3\right) \,,
\eea
and we note that $\dot V^g_y,\dot V^g_z$ are linear in $\mathbf{\dot y,\dot z}$, respectively, i.e.
\ben \label{linear}
\lambda \dot V_y^g (\mathbf{y,\dot y})  = \dot V_y^g (\mathbf{y},\lambda \mathbf{\dot y})\,,
\een
and the same holds for
$\dot V_z^g $. 
Now we substitute the collisional  equations (\ref{y_box3D})-(\ref{z_box3D}) into (\ref{voly})-(\ref{volz}) to obtain
the geometric volume evolution specifically for collisional abrasion
\bea \label{voly2}
\dot V_y^{g,c} (\mathbf{y,\dot y}) & = & \dot V_y^{g,c} (\mathbf{y},a\mathbf{F}^c(\mathbf{y,z})) = a{F}^{g,c}(\mathbf{y,z})\\
\label{volz2}
\dot V_z^{g,c} (\mathbf{z,\dot z}) & = & \dot V_z^{g,c} (\mathbf{z},a\mathbf{F}^c(\mathbf{z,y})) = a{F}^{g,c}(\mathbf{z,y}).
\eea
The geometric volume evolution under friction can be derived similarly to its collisional counterpart in (\ref{voly2})-(\ref{volz2}):
\bea \label{voly2_f}
\dot V_y^{g,f} (\mathbf{y,\dot y}) & = & \dot V_y^{g,f} (\mathbf{y},\mathbf{F}^f(\mathbf{y}, \nu_s, \nu_r)) = {F}^{g,f}(\mathbf{y},\nu_s, \nu_r)\\
\label{volz2_f}
\dot V_z^{g,f} (\mathbf{z,\dot z}) & = & \dot V_z^{g,f} (\mathbf{z},\mathbf{F}^f(\mathbf{z}, \nu_s, \nu_r)) = {F}^{g,f}(\mathbf{z},\nu_s, \nu_r).
\eea
where $\mathbf{F}^f$  is from (\ref{yy_box3D_friction}). We can also compute ${F}^{g,f}(\mathbf{y},\nu_s, \nu_r)$ explicitly
by substituting (\ref{yy_box3D_friction})-(\ref{f_components}) into (\ref{voly}):
\ben \label{fgf}
{F}^{g,f}(\mathbf{y},\nu_s, \nu_r)= \dot V^{g,f}_y=\frac{f^f_1}{y_3}y_2y_3^3+\frac{f^f_2}{y_3}y_1y_3^3+3f^f_3y_1y_2y_3^2=-\frac{V_y}{y_3}(\nu_s+2\nu_r)
\een
where
\bea \label{ff1}
f^f_1(y_1,y_2,\nu_1,\nu_2) & = & \nu_s F^S_1 +\nu_rF^R_1=-\nu_sy_1-\nu_ry_1 \\
\label{ff2}
f^f_2(y_1,y_2,\nu_1,\nu_2) & = & \nu_s F^S_2 +\nu_rF^R_2=-\nu_ry_2 \\
\label{ff3}
f^f_3(y_1,y_2,\nu_1,\nu_2) & = & \nu_s F^S_3 +\nu_rF^R_3= 0.
\eea

\section{Volume weighted individual and mutual abrasion} \label{s:volumeweight}

\subsection{Volume weighted Bloore Equations}
 
Bloore's general equation (\ref{UrBloore}) 
and its particular case (\ref{Bloore}) 
are local in character and did not take into account the possibility
that non-local properties of the pebble might influence
the speed of abrasion $v$. In fact,   
three years before Bloore, Firey \cite{Firey} had studied
a modification of  the Gauss flow (\ref{Gauss}) of the form
\ben
-v= \alpha V^p K \,, \label{Firey} 
\een
where $V$ is the volume of the pebble and $\alpha$ and $p$ are constants.
Based on some experimental work \cite{Archard} 
consistent with the intuition that more massive  pebbles should abrade faster than less massive  particles,
Firey chose $p=1$.  More generally one might consider
replacing  (\ref{Bloore}) by 
\ben
-v= f(V)(1+2bH +cK) \label{volBloore}  
\een
where $f(V)$ may be considered as a variable
speed of attrition for the Eikonal term  depending on the mass 
of equivalently the volume $V$ of the pebble.
We can introduce the volume weight functions in the Binary Bloore flows
(\ref{BinaryBloore1})-(\ref{BinaryBloore2}) as:
\bea
\label{BinaryBloore1_V}
-v_y&=& f^c(V_y,V_z) (1+2\frac{M_z}{4\pi} H_y  +\frac{A_z}{4 \pi}K_y)\\
\label{BinaryBloore2_V}
-v_z&=& f^c(V_z,V_y) (1+2\frac{M_y}{4\pi} H_z  +\frac{A_y}{4 \pi}K_z)
\eea
and in case of spherical particles, based on (\ref{sphere1})-(\ref{sphere2}), this reduces to
\bea \label{mutual1}
-\dot R_y&=& f^c(V_y,V_z)  \left(1+2 \frac{R_z}{R_y}  +\left(\frac{R_z}{R_y}\right) ^2 \right)                        \\
\label{mutual2}
-\dot R_z&=& f^c(V_z,V_y) \left(1+2 \frac{R_y}{R_z}  +\left(\frac{R_y} {R_z}\right) ^2 \right).
\eea

In case of both collisional and frictional abrasion we have

\bea
-v_y&=& f^c(V_y,V_z) (1+2\frac{M_z}{4\pi} H_y  +\frac{A_z}{4 \pi}K_y)
+ f^f(V_y) G(R_y, R_{y\, {\rm min}} , R_{y\, {\rm max}} )  
\\
-v_z&=& f^c(V_z,V_y) (1+2\frac{M_y}{4\pi} H_z  +\frac{A_y}{4 \pi}K_z)
+ f^f(V_z)G(R_z, R_{z\, {\rm min}} , R_{z\, {\rm max} } ) \,,
\label{FricationalBloore_V}\eea

\subsection{Volume weighted Box Equations}

In the box equation approximation one has $V=V(\mathbf{y})=8y_1y_2y_3^3$ 
and (\ref{i_individual_box3D}) becomes
\ben
\label{volBoxeqn}
\dot{\mathbf{y}}=f(V(\mathbf{y}))\mathbf{F}(\mathbf{y},b,c)\,.
\een
Evidently, the path  pursued  by a  pebble in the space of shapes
is unaffected  by the prefactor $f(V)$ in  (\ref{volBloore})
merely the speed with  which the curve is executed. 

We can introduce the volume weight functions in the Binary Box flows (\ref{y_box3D})-(\ref{z_box3D}) as:
\bea
\label{y_box3D_V}
\dot{\mathbf{y}}&=&f^c(V_y(\mathbf{y}),V_z(\mathbf{z}))\mathbf{F}^c(\mathbf{y,z})= f^c(\mathbf{y},\mathbf{z})\mathbf{F}^c(\mathbf{y,z})=
\mathbf{\hat F}^c(\mathbf{y},\mathbf{z})\\
\label{z_box3D_V}
\dot{\mathbf{z}}&=&f^c(V_z(\mathbf{z}),V_y(\mathbf{y}))\mathbf{F}^c(\mathbf{z,y})= f^c(\mathbf{z},\mathbf{y})\mathbf{F}^c(\mathbf{z,y})=
\mathbf{\hat F}^c(\mathbf{z},\mathbf{y})
\eea
where  $\hat{}$  indicates that
the volume weight is included in the operator. The linear behaviour (\ref{linear}) and equations (\ref{voly2})-(\ref{volz2}) imply that in the volume weighted  box equations (\ref{y_box3D_V})-(\ref{z_box3D_V})
volume evolution will be given by
\bea \label{voly2_weighted}
\dot{\hat V}_y^{g,c} (\mathbf{y,\dot y}) & = & \dot{\hat V}_y^{g,c} (\mathbf{y},f^c(\mathbf{y},\mathbf{z})\mathbf{F}^c(\mathbf{y,z})) = 
f^c(\mathbf{y},\mathbf{z}) {F}^{g,c}(\mathbf{y,z})\\
\label{volz2_weighted}
\dot{\hat V}_z^{g,c} (\mathbf{z,\dot z}) & = & \dot{\hat V}_z^{g,c} (\mathbf{z},f^c(\mathbf{z},\mathbf{y})\mathbf{F}^c(\mathbf{z,y})) =
f^c(\mathbf{z},\mathbf{y}){F}^{g,c}(\mathbf{z,y})
\eea
where $\hat{} $ refers to the inclusion of the volume weight function  and ${F}^{g,c}$ is given in (\ref{voly2}).
We introduce the volume weight function in an analogous manner for frictional abrasion based on (\ref{yy_box3D_friction}):
\ben 
\label{yy_box3D_friction_V}
\dot \mathbf{y}  =  f^f(V_y(\mathbf{y}))\mathbf{F}^f(\mathbf{y},\nu_s,\nu_r) = f^f(\mathbf{y})\mathbf{F}^f(\mathbf{y},\nu_s,\nu_r) = \mathbf{\hat F}^f(\mathbf{y},\nu_s,\nu_r)
\een
and again $\hat{}$  indicates that
the volume weight is included in the operator. Here again (\ref{linear}) and (\ref{voly2_f})-(\ref{volz2_f}) imply that in  volume weighted frictional box equation (\ref{yy_box3D_friction_V})
volume evolution is given by:
\ben \label{voly2_weighted_friction}
\dot {\hat {V}}_y^{g,f} (\mathbf{y,\dot y})  = \dot {\hat {V}}_y^{g,f} (\mathbf{y},f^f(\mathbf{y})\mathbf{F}^f(\mathbf{y},\nu_s,\nu_r)) =
f^f(\mathbf{y}) {F}^{g,f}(\mathbf{y},\nu_s, \nu_r)
\een
where ${F}^{g,f}$ is given in (\ref{fgf}) and $\dot {\hat{V}}_z^{g,f} (\mathbf{z,\dot z})$ is defined in the same manner.

Our next goal is to derive the volume weight function $f(V_y,V_z)$ for the Binary Bloore Flows 
(\ref{BinaryBloore1_V})-(\ref{BinaryBloore2_V}) and Binary Box Flows (\ref{y_box3D_V})-(\ref{z_box3D_V}), based on some physical considerations
and to investigate the stability of the volume-weighted self-dual flows. The PDE (\ref{BinaryBloore1_V})-(\ref{BinaryBloore2_V}) only admits
the study of the special case where
both particles are spherical and we will derive the volume weight function for this case. Subsequently, in an analogous manner,
we will identify the volume weight function for general (non-spherical) particle evolution in the box equations (\ref{y_box3D_V})-(\ref{z_box3D_V}).

\subsection{Asymmetry of the volume weight function stabilising the self-dual flows}
Before introducing the physical considerations, we point out, purely on geometric grounds, a fundamental property of the volume weight function $f$:
in order to stabilise the self-dual collisional flows, $f$ \emph{needs to be} asymmetrical. It is sufficient to show in the spherical case
that the symmetric volume weight
function implies instability.

The spherical flow (\ref{mutual1})-(\ref{mutual2}) takes place in the positive quadrant of the
$R_y-R_z$ plane with both $R_y$ and $R_z$ decreasing. 
Defining, as is standard 
\ben
\tan \theta = \frac{R_z}{R_y}\,,\qquad \tan \psi = \frac{dR_z}{dR_y}
\een
we find that the trajectories satisfy 
\ben
\frac{dR_z}{dR_y} = \frac{f(V_z,V_y)}{f(V_y,V_z)} \cot ^2\theta \,, 
\een
or in terms of volumes:
\ben
\frac{dV_z}{dV_y} = \frac{f(V_z,V_y)}{f(V_y,V_z)}. \label{volumeflow}
\een
It is immediately apparent that if $f$ is symmetric, i.e. 
\ben
f(V_z,V_y)= f(V_z,V_z) 
\een
then  we have \ben
\frac{dV_z}{dV_y} = 1\,,
\een
that is the trajectories are straight lines in the $V_y,V_z$ plane
making an angle of $\frac{\pi}{4}$ with the axes. 
By using  $V_y=\frac{4 \pi}{3} R_y^3$ and  $V_z=\frac{4 \pi}{3} R_z^3$, these can be transferred to   the $[R_z,R_y]$ plane where straight lines become 
curves which in the downward direction move away from the straight line $R_z=R_y$.
It follows that if the volume weight function $f(V_y,V_z)$ is symmetrical then the self-dual trajectory  defined by $R_y=R_z$ 
is  unstable within the class of spherical flows. Beyond showing that asymmetry is a necessary condition for the stability
for the self-dual flows, we also show a simple example where it is also sufficient. If we assume that 
\ben \label{asym1}
f(V_y,V_z) = \left( \frac{V_y}{V_z}\right) ^p 
\een
then we have
\ben
\frac{dR_z}{dR_z} = \tan \psi  =  (\tan \theta) ^{2(3p-1)}. \label{slope}  
\een
If $p<\frac{1}{3}$
and the trajectory   lies above the diagonal line $\theta = \frac{\pi}{4}$,
then its slope $\psi$ is less than  $\frac{\pi}{4}$ and it will move away from
the diagonal. If the trajectory   lies below  the diagonal then its 
slope $\psi$ is greater  than  $\frac{\pi}{4}$ and it will again move away from
the diagonal. Thus if $p \le \frac{1}{3}$ the self-dual flow is unstable
and if  $p > \frac{1}{3}$ it will be stable.   

As pointed out in \cite{DG}, friction  can stabilize attractors in the geometric self-dual flows in the $[y_1,y_2]$ space of box ratios.
Here we would like to point out that in case of volume-weighted spherical flows, friction also contributes to the relative stabilisation of
size in the sense that the particle's linear size converges to each other. Since we treat friction as an individual abrasion, any monotonically increasing volume weight function $f^f(V_y)$ associated with friction produces
an asymmetry which has an analogous effect to the above-discussed asymmetry of the volume weight function
for collisional abrasion.

In the next section we show
that asymmetric models (although more complex than (\ref{asym1})) emerge naturally from physical considerations.
We will only prove  the stabilising property of the physical volume weight functions for the spherical case, however,
they appear to have the same effect for general geometries.

\subsection{Derivation of the volume weight function from physical models in the Bloore equations}

We assume that volume evolution is given by an independent physical model as

\bea
\dot V^p_y  & = & C^c_yg^c(V_y,V_z) \\
\dot V^p_z & = & C^c_zg^c(V_z,V_y).
\eea
where the superscript $p$ stands for "physical" and the constants $ C^c_y,C^c_z$ may differ due to the different hardness of 
the material of the particles.In the spherical flows we can use (\ref{volumesphere}) to obtain the volume weight function as
\ben \label{sphereweight1}
f(V_y,V_z)=\frac{C^c_yg^c(V_y,V_z)}{4a\pi(R_y +R_z)^2}.
\een
Using (\ref{sphereweight1}),
(\ref{mutual1})-(\ref{mutual2}) can be written as
\bea \label{mutual1_weight}
-\dot R_y&=& \frac{C^c_yg^c(V_y,V_z)}{4a\pi(R_y +R_z)^2} \left(1+2 \frac{R_z}{R_y}  +\left(\frac{R_z}{R_y}\right) ^2 \right)\\
\label{mutual2_weight }
-\dot R_z&=& \frac{C^c_zg^c(V_z,V_y)}{4a\pi(R_y +R_z)^2} \left(1+2 \frac{R_y}{R_z}  +\left(\frac{R_y} {R_z}\right) ^2 \right).
\eea
Later we give examples for some specific functions $g^c(V_y,V_z)$.

\subsection{Derivation of the volume weight function from physical models in the box equations}

Without giving any specific physical abrasion model, in this subsection we show how the volume weight functions $f^c,f^f$ can be formally derived
if such models are available. Later on, we give specific examples of some physical models, however, any physical model can be plugged into the equations
of this subsection. We only assume that the physical model is defined by  volume evolution equations for collisional and frictional abrasion, respectively, as

\bea \label{phys1}
\dot V_y^{p,c} & = & C^c_yg^c(\mathbf{y},\mathbf{z})\, \qquad  \dot V_y^{p,f}  =C^f_yg^f(\mathbf{y}) \\
\label{phys2}
\dot V_z^{p,c} & = & C^c_zg^c(\mathbf{z},\mathbf{y})\, \qquad \dot V_z^{p,f}  =C^f_zg^f(\mathbf{z}),
\eea
then by using (\ref{voly2_weighted})-(\ref{volz2_weighted}) and (\ref{voly2_weighted_friction}) we can set the geometric and physical volume evolution rates
to be equal and this condition yields:

\bea \label{volumeweight_c}
f^c(\mathbf{y},\mathbf{z}) & = & \frac{C^c_yg^c(\mathbf{y},\mathbf{z})}{{F}^{g,c}(\mathbf{y,z})}  \\
\label{volumeweight_f}
f^f(\mathbf{y}) & = & \frac{C^f_yg^f(\mathbf{y})}{{F}^{g,f}(\mathbf{y})}
\eea
and  ${F}^{g,c}$ and ${F}^{g,f}$ are given in (\ref{voly2}) and (\ref{voly2_f}), respectively. So, based on 
the above equations and (\ref{y_box3D_V})-(\ref{z_box3D_V}) and (\ref{yy_box3D_friction_V}), the box equations for the combined model (including the physical law for volume evolution) are
\bea \label{unified_y}
\dot \mathbf{y} & = & \frac{C^c_yg^c(\mathbf{y},\mathbf{z})}{{F}^{g,c}(\mathbf{y,z})} \mathbf{F}^{c}(\mathbf{y,z})+ \frac{C^f_yg^f(\mathbf{y})}{{F}^{g,f}(\mathbf{y})}
\mathbf{F}^{f}(\mathbf{y},\nu_s,\nu_r)=\mathbf{F}^u(\mathbf{y,z})\\
\label{unified_z}
\dot \mathbf{z} & = & \frac{C^c_zg^c(\mathbf{z},\mathbf{y})}{{F}^{g,c}(\mathbf{z,y})} \mathbf{F}^{c}(\mathbf{z,y})+ \frac{C^f_zg^f(\mathbf{z})}{{F}^{g,f}(\mathbf{z})}
\mathbf{F}^{f}(\mathbf{z},\nu_s,\nu_r) = \mathbf{F}^u(\mathbf{z,y}),
\eea
where $\mathbf{F}^{c}, \mathbf{F}^{f}$ are defined  in (\ref{i_individual_box3D}), (\ref{y_box3D}) and (\ref{yy_box3D_friction}), respectively
and ${F}^{g,c}$,${F}^{g,f}$ are given in (\ref{voly2}),(\ref{voly2_f}).

\section{Collective abrasion}\label{s:collective}

Using the above model, a Markov-process can be simulated by regarding 
 $\mathbf{y},\mathbf{z}$ in (\ref{unified_y})-(\ref{unified_z}) as random vectors with \em identical \rm
distributions since they represent two random samples of the same pebble population. 
The evolution of this Markov process (and thus the time evolution of
of the pebble size and ratio distributions) is of prime interest since it determines the physical relevance of the stable
attractors identified in \cite{DG}.
While the analytical investigation of the Markov
process is beyond the scope of this paper, direct simulations are relatively straightforward.
We consider $N$ pebbles out of which we randomly draw two with coordinates
$\textbf{y}^0,\textbf{z}^0$ and run equations (\ref{unified_y})-(\ref{unified_z}) for a short time period
$\Delta t$ on these initial conditions to obtain the
updated vectors $\textbf{y}^1$,$\textbf{z}^{1}$ . In the simplest linear approximation we have the recursive formula
\bea
\label{Nbody1} 
\textbf{y}^{i+1} & = & \textbf{y}^{i} + \Delta t \textbf{F}^u(\textbf{y}^i,\textbf{z}^i)\\
\label{Nbody2}
\textbf{z}^{i+1} & = & \textbf{z}^{i} + \Delta t \textbf{F}^u(\textbf{z}^i,\textbf{y}^i)\,.
\eea
Such an iterative step can be regarded as the cumulative, averaged effect of several collisions between
the two selected pebbles.  Apparently, the $N=2, \Delta t \to 0$ case is 
identical to (\ref{unified_y})-(\ref{unified_z}).
In \cite{DG} we investigated the behaviour of the deterministic flows in the
special cases of steady state flows (\ref{yy_box3D})-(\ref{y3_box3D}) and self-dual flows (\ref{Selfdual_box}).
Multi-body simulations allow 
 the numerical study of the statistical stability of the
flows, i.e. one can assess the stability of the above-mentioned special cases.

\section{Physical models of mass evolution} \label{s:phys}

It appears to be  widely believed that the relationship between
volume $V$ and time  $t$    follows an exponential law 
suggested by Sternberg \cite{Sternberg}
\ben
V(t) = V(0) e^{-\frac{t}{t_0} }\,, \label{Sternberg}
\een
where $t_0$ is a constant. More accurately, Sternberg's Law is 
usually held to hold for the volume  of pebbles as a function of distance
along a river or stream. If they are transported along the river at constant
speed  this is equivalent to (\ref{Sternberg}).

Our goal is to introduce physical collisional models which, on one hand, predict infinite lifetimes (in accordance with Sternberg), on the
other hand, they can be plugged into the geometric equations via the formulae  (\ref{unified_y})-(\ref{unified_z}). We propose first collisional 
models followed by frictional models.

\subsection{Collisional Models }

It seems intuitively reasonable 
that that mutual abrasion will be greater
the greater the  kinetic energy $E_{\rm com} $ of the colliding particles 
in their common rest frame. This is given by
\ben \label{energy}
E_{\rm com}= \half \frac{ m_ym_z}{m_y+m_z} u^2
\een
where $u$ is the relative velocity of the abrader and the abraded
and $m_y$ and $m_z$ are the masses of the pebbles.
These will be related to the densities $\rho_y$ and $\rho_z$ 
and volumes by 
by
\ben
m_y= \rho_y V_y \,, \qquad m_z= \rho_z V_z \,.
\een
For a homogeneous ensemble of pebbles it is reasonable to assume
$\rho_y=\rho_z$. In binary collisions one might suppose that the rate of reduction of volume
is proportional to $E_{com}$
and a power $\alpha$ of the mass. Assuming equal densities 
and that $u^2$ is on average a constant, we arrive at the equation for \emph{physical volume evolution}
\bea \label{volevoly}
-\dot V^{c,p}_y & = & C^c_y V_y^\alpha \frac{V_yV_z}{V_y+V_z}=C^c_yg^c(V_y,V_z)\\
\label{volevolz}
-\dot V^{c,p}_z & = & C^c_z V_z^\alpha \frac{V_yV_z}{V_y+V_z}=C^c_zg^c(V_z,V_y)
\eea
where the superscript $p$ stands for "physical" and the constants $ C^c_y,C^c_z$ may differ due to the different hardness of 
the material of the particles.
This results in
\ben \label{volevol}
\frac{dV_z}{dV_z} = \frac{C^c_z}{C^c_y}\left(\frac{V_z }{V_y} \right)^\alpha .
\een
We remark that one plausible motivation behind (\ref{volevoly})-(\ref{volevolz}) is Weibull Theory for fragmentation
 \cite{Weibull} \cite{Kun1} relating the material strength $\sigma _{crit}$ to the specimen
mass $m$ as
\ben \label{Weibull}
\sigma_{crit}=\sigma_0\left(\frac{m}{m_0}\right)^{-\frac{1}{\mu}}
\een
where $\sigma_0$ is the strength of the specimen of unit volume $m_0$
and $\mu$ is Weibull's modulus. This formula is based
on the statistical distribution of Griffith cracks
\cite{Griffith} and $\mu \to \infty$ corresponds to homogeneous
material without Griffith cracks. Here we assume that the critical
fragmentation energy $E_f$  per fragmented mass $m_f$, given as
\ben \label{Weibullenergy}
\tau_{crit}=\frac{E_f}{m_f}
\een
follows a similar power law
\ben \label{strength}
\tau _{crit}=\tau_0\left(\frac{m}{m_0}\right)^{-\frac{1}{\bar\mu}}
\een
and similarly to Weibull's modulus, $\bar \mu\to \infty$ corresponds  to homogeneous
material.  Using equations (\ref{energy}),(\ref{Weibull}) and (\ref{Weibullenergy})  yields (\ref{volevoly})-(\ref{volevolz})
with $\alpha = 1/\bar \mu$. 
Note that $\alpha=0$ corresponds to homogeneous material.
As pointed out in \cite{Kun1}, brittle materials are \emph{softening} in fragmentation in
the sense that the energy per unit fragmented volume is decreasing with the size of the particle.
This behaviour implies in (\ref{volevoly})-(\ref{volevolz})
\ben \label{softening}
\alpha \ge 0.
\een
In the box equations, via (\ref{vol1y})-(\ref{vol1z}), (\ref{volevoly})-(\ref{volevolz}) is translated into
\bea \label{volevoly_box}
-\dot V^{c,p}_y & = &C^c_yg^c(\by,\mathbf{z})\\
\label{volevolz_box}
-\dot V^{c,p}_z & = &C^c_zg^c(\mathbf{z}, \by)
\eea
which can be plugged into (\ref{unified_y})-(\ref{unified_z}). In the spherical case we have
\ben \label{sphere_physical}
g^c(R_y,R_z)=\left(\frac{4\pi}{3}\right)^{1+\alpha}\frac{R_y^{3(1+\alpha})R_z^3}{R_y^3+R_z^3}
\een
and using (\ref{sphereweight}) this yields for the volume weight function
\ben \label{sphereweight}
f(V_y,V_z) = \frac{C^c_y}{3a} \left(\frac{4 \pi}{3}\right)^\alpha 
\frac{R_y^3R_z^3}{(R_y^3+R_z^3)}  \frac{R_y^{3 \alpha}} {(R_y+ R_ z )^2} 
\een
By substituting (\ref{sphere_physical}) into (\ref{mutual1_weight}) we get the physical evolution equations for spheres.

We also note that (\ref{sphereweight}) is asymmetrical: $f(V_y,V_z) \ne f(V_z,V_y)$.
Indeed, in the case of spheres, (\ref{volevoly})-(\ref{volevolz}) yield
\ben
\frac{dR_z}{dR_z} = \left(\frac{R_z }{R_y} \right)^{3\alpha -2}  
\een
and we can immediately see that the self-dual trajectory $R_y=R_z$ will therefore
be unstable unless $\alpha > \frac{2}{3}$. Recalling that the exponent $\alpha$ was motivated by Weibull
theory, this condition suggests that, in the absence of other effects, for nearly homogeneous particles
the self-dual flows will be unstable.

\subsection{Frictional models}

Here we describe the evolution of mass as a single particle $K_y$ is being abraded by friction
and we postulate
\ben \label{myf}
-\dot m_y =  \bar C^f_y m_y^{\beta}, \qquad \bar C^f_y>0 \\
\een
which, for $\beta=1$ is essentially a simplified version of Archard's formula \cite{Archard} by assuming constant velocity and contact area with the abrading surface.
If the contact stress approaches the yield stress then higher $\beta$ values may be appropriate.
The case $\beta \geq 1$ corresponds to infinite time horizon and, as we will show in the next subsection, the volume evolution equations (\ref{volevoly})-(\ref{volevolz})
also predict similar behaviour, so for $\beta \geq 1$  the two effects (collisional and
frictional abrasion) may compete on the same timescale. 
In equation (\ref{myf}), $\beta \geq 1$ can be motivated by assuming friction caused entirely by the gravity acting on the particle $K_y$,
e.g. the particle is sliding on a free surface. Friction could also occur inside granular assemblys under compressive
forces far exceeding the particles own weight; in this case mass will decay in finite time and frictional
abrasion will dominate the whole process. However, as we showed in \cite{DG}, only the continuous interaction
of collisional and frictional abrasion can produce the geologically observed dominant pebble box ratios.
Based on (\ref{myf}) we have
\ben \label{vyf}
-\dot V^{p,f}_y  = C^f_y V_y^{\beta}=C^f_yg^f(V_y) 
\een
where $C^f_y=\bar C^f_y/\rho_y$ and again, the superscript $p$ refers to the fact that this evolution is based
on physical considerations rather than geometrical ones, superscript $f$ refers to the frictional process.
In the box equations (\ref{vyf}) translates into
\ben \label{vyf_box}
-\dot V^{p,f}_y  = C^f_y V_y^{\beta}=C^f_y(8y_1y_2y_3^3)^{\beta}=C^f_yg^f(\by)
\een
which can be plugged into (\ref{unified_y}).

\subsection{Collective abrasion: rescaling of time}

In section \ref{s:collective} we introduced the concept of collective abrasion.
In case of  two particles under mutual collisions we have assumed that in equal time intervals equal number
of collisions occur.  If we consider a collection of particles from which we choose random pairs
and evolve them under the above-described binary process then the choice of this
pairs can follow various rules, in any case, we have to consider that the probability of 
collision in equal time between two arbitrary particles is not equal.
For example, it is a plausible assumption that in the same amount of time a large particle will suffer more
collisions than a small particle. We will implement the particular assumption
that the number $N_y$ of collisions per unit time suffered by the particle
$\by$ is proportional to the $\nu$-power of the relative volumes:
\ben
N_y\propto \left(\frac{V_y}{V_z}\right)^{\nu}.
\een
Needless to say, this assumption would not make sense in the binary process since
from it would follow that the two colliding particles suffer different number
of collisions in equal time intervals. Nevertheless,  in case of collective
abrasion this assumption can be implemented and in essence it requires
the rescaling of time. If we denote the time in the collective process by $T$
and time in the original, binary process by $t$ then we have
\ben
\frac{dT}{dt} = \left(\frac{V_y}{V_z}\right)^{\nu}.
\een
If we study the collective process (\ref{Nbody1})-(\ref{Nbody2}) process 
then rescaled time can be implemented by modifying (\ref{volevoly})-(\ref{volevolz}) as

\bea \label{volevoly_c}
-\dot V^{c,p}_y & = & C^c_y\frac{V_y^{(\alpha+\nu +1)}V_z^{(1-\nu)}}{V_y+V_z}=C^c_y\bar g^c(V_y,V_z)= C^c_y\bar g^c(\by,\mathbf{z})\\
\label{volevolz_c}
-\dot V^{c,p}_z & = & C^c_z\frac{V_z^{(\alpha+\nu +1)}V_y^{(1-\nu)}}{V_y+V_z}=C^c_z\bar g^c(V_z,V_y)= C^c_z\bar g^c(\mathbf{z},\by).
\eea
As a consequence, if we model collective abrasion then  in (\ref{unified_y})-(\ref{unified_z})  $g^c(\by,\mathbf{z})$ has to be replaced by $\bar g^c(\by,\mathbf{z})$ and
all other formulae remain unchanged.

\section{Lifetimes, Sternberg's Law and the stability of the self-dual flows} \label{s:stab}

\subsection{Lifetimes, physical mass evolutoion models and Sternberg's Law}

Bloore's geometric equation apparently predicts finite lifetimes for abrading particles,
this is immediately suggested by the constant term on the right hand side of (\ref{Bloore}).
However, not only the constant, but every single term in the geometric equation predicts
finite time horizon for the particle and this property is inherited by the box equations, we discuss this in Appendix \ref{app:box}.

Our box model (\ref{unified_y})-(\ref{unified_z}) is constructed in such a way
that geometric volume evolution rates ${F}^{g,c}$,${F}^{g,f}$  (given in (\ref{voly2}),(\ref{voly2_f}))
are completely suppressed and volume evolution is determined by the physical evolution rates
given in (\ref{phys1})-(\ref{phys2}). Consequently, the lifetimes for the unified
box model (\ref{unified_y})-(\ref{unified_z}) are determined by the lifetimes
for the physical volume evolution models (\ref{phys1})-(\ref{phys2}) and next we study the latter.
As we are about to show, they predict exponential decay for the volume, thus reproducing
the empirical law (\ref{Sternberg}) of Sternberg \cite{Sternberg}. Needless to say, these
models are certainly not unique and others may have similar properties.

We gave two examples of physical evolution models for collisional abrasion in (\ref{volevoly})-(\ref{volevolz})
and (\ref{volevoly_c})-(\ref{volevolz_c}). Since the former is just the $\nu=0$ special case of the latter
it suffices to study the latter. We introduce a simple 

\medskip

\noindent {\bf Lemma}
{\it The differential equation 
$\dot f = -c f^\gamma$  (with $c={\rm constant}>0, f(t_0)>0, \gamma \ne 1)$  has a solution  
$f(t) = \Bigl ( f^{1-\gamma} (t_0) - (1-\gamma ) (t-t_0) \Bigr )^
{1/(1-\gamma)}$ for $t\ge t_0$.
Thus if $\gamma <1$, $f(t)$ goes to zero in finite time, whereas if $\gamma >1$,
then  $f(t) $   reaches zero only after an infinite time.}

Similar conclusions could be reach if $c(t)$ varies with time,
with $c(t-t_0) $ replaced by $\int_{t_0} ^t c(t) dt$. In particular, if $c(t) \to 0$
and $\gamma >1$ then we also have infinite time horizon.
Equation (\ref{vyf}) describes mass and volume evolution under friction,
trivially agree with the equation in the Lemma and for $\beta >1$  it corresponds to processes with infinite
lifetimes. 
Next we consider equations (\ref{volevoly_c})-(\ref{volevolz_c}) for mass evolution under collisional abrasion. 
We note that in (\ref{volevoly_c})-(\ref{volevolz_c}) both variables are strictly monotonically
decreasing, regardless of the initial values. This implies that two cases
are possible: (I)  either $V_y$ or $V_z$ will approach zero while the other volume
is still finite or
(II) when  both volumes approach zero simultaneously at some slope $V_z/V_y=c_0$.

\medskip
\noindent{\bf Case (I)}  Assume $V_y$ approaches zero first and thus we have $Vy << V_z$. Then, if $\nu=0$,
equation (\ref{volevoly_c}) for $\dot V_y$ may be approximated by the equation  
in the lemma, by setting $f= V_y$, $\gamma = \alpha+1$ , $c=- C^c_y$.
By assumption (\ref{softening}), $\alpha \ge 0$ and so
in all cases $\gamma \ge 1$. It follows that
the lifetime for the $\by$ particle  is always infinite, approaching $V_y=0$
asymptotically. As $V_y$ is asymptotic to zero,
based on (\ref{volevolz_c}) so is $\dot V_z$, so the $\mathbf{z}$ particle will also have
infinite time horizon (approaching finite constant mass). 
If $\nu \geq 0$ then we have $c(t)=(V_y/V_z)^\nu$ and since $V_y \to 0$ we also have $c(t) \to 0$
so this also yields infinite time horizon for both particles.

\medskip 
\noindent {\bf Case (II) }  If $V_y$ and $V_z$  vanish together at some slope $V_z/V_y=c_0$
then we can take either
to equal $f$ in the lemma and $\gamma =\alpha+1$, $c=-c_0 C^c_y/(c_0+1)$ or $c=-c_0C^c_z/(c_0+1)$.
The same conclusion holds.

\subsection{Lifetimes and the  volume weight functions}
Field observations of river pebbles  
are consistent with Sternberg's Law  \cite{Sternberg} which predicts 
that particles live for ever.  This gives an important constraint
on evolution laws. In our model, the latter determine the volume weight functions and next
we shall give some general results
on whether or not a volume weight function predicts a finite lifetime
by giving a general upper bound on the lifetime. 

In the spherical case, based on (\ref{mutual1}) we can write
\ben \label{ry}
-\dot R_y \ge f(V_y,V_z) 
\een
and so we have
\ben
\frac{R_y(t)}{R_y(0)} \le e^{ - \int _0^t \frac{f(V_y,V_z)} {R_y} dt ^\prime
}\een
which gives exponential decay as long as 
$\frac{f(V_y,V_z)}{R_y}$ converges for small $R_y$.
In the general case we may obtain volume evolution by integrating (\ref{ry}) over the surface $\Sigma$:
\ben
-\dot V_y \ge \int_\Sigma f(V_y,V_z) dA =A_yf(V_y,V_z).
\een 
Thus we have
\ben
\frac{V_y(t)}{V_y(0)} \le e^{-\int _0^t 
f(V_y,V_z) \frac{A_y }{V_y}  dt ^\prime }   
\een
which gives exponential decay as long as 
$f(V_y,V_z) \frac{A_y }{V_y}$ converges for small $V_y$.

\subsection{Stability of the self-dual flows in the stochastic process}
We can study the evolution of $\rho=V_y/V_z$ under the described process and we can see
that $\rho=1$ is always a solution of (\ref{volevoly_c})-(\ref{volevolz_c}).
The stability of this solution is of particular interest since it indicates
the stability of the self-dual flows in  (\ref{unified_y})-(\ref{unified_z}).
It is easy to see that the stability of $\rho=1$ is guaranteed if
\ben \label{stab}
(1-\rho)\dot\rho>1
\een
and we can see from (\ref{volevoly_c})-(\ref{volevolz_c}) that the condition for stability is
\ben
\alpha + 2\nu > 1.
\een

\begin{figure}[ht]
\begin{center}
\includegraphics[width=75 mm]{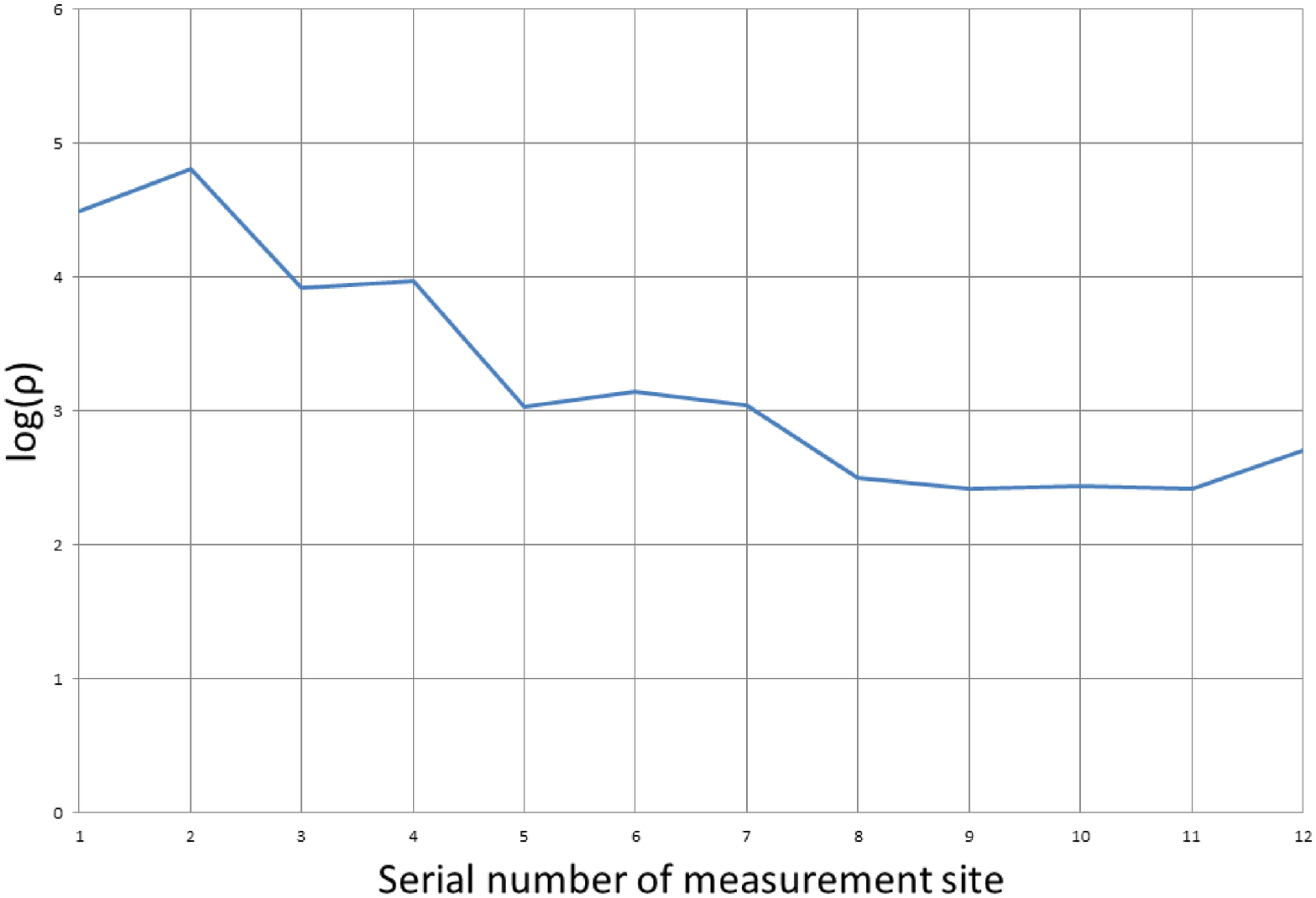}
\end{center}
\caption{Field data from the Williams river}\label{Fig1}
\end{figure}

\noindent
Now, we expect  $\alpha<<1$ if the material is nearly homogeneous; this suggests that
the self-dual flows are not stable in the binary process where $\nu=0$. In other words, our model predicts that the mass
ratio of two, mutually abrading particles will diverge if the material
is sufficiently homogeneous. On the other hand, $\nu=2/3$ is a plausible assumption,
relating the number of collisions per unit time to the effective cross section
of the particle. So, in a collective process we expect that the self-dual flows
will be stable and attractive. This is also confirmed by the field
data collected along the Williams river where we measured
$\bar \rho=V_{max}/V_{min}$ in each sample. Since
$\bar \rho$ is an upper bound for $\rho$, its evolution indicates
the stability of the $\rho=1$ solution. In Figure \ref{Fig1} we
plotted $\log(\bar \rho)$ versus the serial number of the measurement
site along the Williams river, the latter can be regarded as an
approximate measure of time. As we can see,  $\log(\bar \rho)$ shows
a marked decrease along the river thus indicating the stability of the
$\rho=1$ solution.

These considerations also imply that our conclusions regarding the
role of segregation in \cite{DG} are only valid for the geometric equations.
If we study the unified flows then we expect that under the combined
effect of collisions and friction, stable attractors in the space
$[y_1,y_2]$ of the box ratios will emerge spontaneously and robustly.
Also, while segregation by size is catalysing this process, it is
not a pre-condition for the emergence of the attractors. Rather,
we expect that abrasion itself will further help to produce
pebbles of similar sizes.

\section{Acknowldegements}
This research was supported by OTKA grant T104601. The comments from Dr Timea Szab\'o are greatly appreciated.

\section{Appendix: Uni-directional Bloore Flows and Weingarten Surfaces}  \label{s:wein}
Bloore
originally proposed \cite{Bloore} his equation to
describe the evolution of the  surface
of a pebble under  {\sl isotropic} 
bombardment by abraders. In the case of  bedrock evolution 
for example, the abraders are {\it unidirectional}  and a modification
of his equation is required. The simplest modification is the insertion
of an inclination factor  $\cos\theta$, where $\theta$ is
the angle between the direction of the abraders and the normal of the abraded
surface\cite{DGS}.  This amounts to replacing 
$v$ by $\frac{v}{\cos \theta}$ in the previous equations.
In a previous paper \cite{DGS} we showed how circular
profiles evolving with constant speed $u$ , sometimes called {\it translators}  
emerge
as stable final states of the cylindrically symmetric (or planar) 
form of the unidirectional Bloore equation. This agreed with
existing   theoretical and experimental work described in \cite{S,DSV}.

In this section we shall  extend our earlier result to the
full three-dimensional case. We find that the possible
final states are in general  Weingarten surfaces, that is \cite{Forsyth}
surfaces
for which there is a functional relation between the two
principal curvatures $\kappa_1$ and $\kappa_2$ .
In the special case  (\ref{Bloore})  
the possible final states belong to   
 a special class of Weingarten surfaces   
(sometimes called {\it linear})\footnote{Beware: Linear Weingarten surfaces are
sometimes defined differently:  such that 
there is a linear relation between the principle curvatures.
This is not equivalent. Another terminology for what we call
Linear Weingarten surfaces is {\it Special Weingarten Surfaces }or
 {\it SW surfaces} . However this use is by no means universal }, 
 whose mean curvature $H$, and Gauss curvature $K$ satisfy
the linear relation
\ben
f(1+ 2b H + c K )  = u \,,\label{line}
\een
where, $f,b,c$ are constants characterising the abraders, and
$u$ is the constant final speed. A possible test of
the theory would be examine the distribution of mean and Gauss curvature   
on an abraded rock face as a function of time.  
If governed by the unidirectional
Bloore equation this distribution, when plotted in the  $K-H$
plane should concentrate on the straight line (\ref{line})

\subsection{Weingarten surfaces as translators}

We choose, for concreteness,  to work with the
Monge representation in which the original Bloore
equation is (\ref{Monge}), but our result does not depend on
that choice.  The cosine $\cos \theta$  
between the normal and the positive $z$ direction
is given by 
\ben
\cos \theta = \frac{1}{\sqrt{1+h_x^2 +h^2_y}}\,.
\een
Replacing $v$ by   $\frac{v}{\cos \theta} $ in (\ref{Monge}) 
gives
\ben
\dot h = v(\kappa_1, \kappa_2)   
\een
 If $u$ is the constant speed, we have 
\ben
h=ut+ z(x,y) \,,
\een
and therefore
\ben
u = v(\kappa_1, \kappa_2)\,,
\een
which shows that the translator must be Weingarten surface.
In the special case (\ref{Bloore}) we
obtain (\ref{line}).   

A simple example of a  travelling front or translator solution is
a sphere 
\ben
h(x,y,t) =ut + \sqrt{R^2-x^2-y^2}\,,\qquad  u= f(1+\frac{b}{R} + \frac{c}{R^2})   \een
If  $c=0$  we obtain {\it surfaces of constant  mean curvature} 
\ben
H= \half \frac{u-f}{fb} 
\een
The case $u-f=0$ gives
\ben
R_1+ R_2 = - \frac{c}{b} 
\een
If $b=0$ then
\ben
fc \frac{1}{R_1R_2} = u-f
\een
which are surfaces of constant curvature.
If $ \frac{1}{4} \Delta = f^2b^2 -(f-u) fc  $   
then if $\Delta >0$ the  linear Weingarten surface is called  
{\it elliptic}, if $\Delta
<0$ it is called {\it hyperbolic} and if $\Delta =0$ it is called
{\it tubular}. In particular, a surface of constant negative
curvature is hyperbolic, while  surfaces of constant  positive curvature
are elliptic, as are surfaces of constant mean curvature.  

An example \cite{Brunt} of a hyperbolic  surface of revolution given by
\ben
(x,y,z) = (\rho(u) \cos \phi, \rho(u) \sin \phi, z(u)) 
\een
with
\ben
\rho(u) = \sin u-\cos u\,,\qquad z(u)= \cos u + \sin u + \ln 
\Bigl( \frac{\sin u }{1+ \cos u}     \Bigr ) 
\een
which satisfies
\ben
H+K+\half =0\,.
\een

\section{Appendix: Observable quantities in the geometric equations } \label{app:box}

We study particle shape evolution under collisional abrasion,
governed by Bloore's partial differential equation (\ref{Bloore})
and we are concerned about qualitative and quantitative features of the evolution of  the \emph{observable quantities} 
such as linear size (maximal width), surface area and volume (denoted by $D(t), A(t), V(t)$, respectively) associated with convex solids in collisional abrasion governed
by Bloore's partial differential equation (\ref{Bloore}).
 We refer to the three terms as the Eikonal,
Mean Curvature and Gaussian term, respectively.  

In  (\ref{Bloore}), if coefficients $b,c$ are constant then 
all observable quantities have finite lifetimes.
If  only one component of (\ref{Bloore}) is acting then we have:
\bea
\label{eikonal}
\mbox{\bf Eikonal:} & \dot D & = -1, \\
\label{mean}
\mbox{\bf Mean Curvature:} & \dot A & = -2b(W+4\pi), \\
\label{gauss}
\mbox{\bf Gaussian:} & \dot V & = -4\pi c.
\eea
where $\dot()$ refers to differentiation with respect to time $t$
and $W$ is the Wilmore functional given by
\ben \label{Wilmore}
W= \int _\Sigma   \frac{1}{4} 
\bigl(\frac{1}{R_1}-\frac{1}{R_2}  \bigr ) ^2      dA  \ge 0 \,
\een
where $R_i$ are the principal radii. In case of the box equations (\ref{y_box3D})-(\ref{z_box3D}) we have the analogous observable quantities
$D_{box},A_{box}, V_{box}$, all given as functions of the dimensionless box
ratios $y_1,y_2$ multiplied by some power of $y_3$:
\bea
\label{dbox}
D_{box} & = & 2y_3 \\
\label{abox}
A_{box} & = & 8y_3^2(y_1y_2+y_1+y_2)\\
\label{vbox}
V_{box} & = & 8y_3^3y_1y_2
\eea
In case of the Eikonal, Mean Curvature and Gaussian flows we have
\bea
\label{eikonalbox}
\mbox{Eikonal} &:& \dot y_3 = -2 \\
\label{meanbox}
\mbox{Mean Curvature}&:&  \dot y_3 = -\frac{1}{y_3}\left(\frac{y_1^2+y_2^2}{2y_1^2y_2^2}\right) \\
\label{gaussbox}
\mbox{Gaussian}&:& \dot y_3 = -\frac{1}{y_3^2}\frac{1}{y_1^2y_2^2}
\eea
so, based on (\ref{dbox})-(\ref{gaussbox}),
the evolution speed for observable quantities is in all cases
a product of an $n$-th power of $y_3$  and some function $f(y_1,y_2)$:
\ben \label{o1}
\frac{d}{dt}\{Observable\}=y_3^nf(y_1,y_2)
\een
The resulting values for $n$ are summarised in Table 1 where we also list in $[]$ brackets
the power of length in the evolved observable quantity and the geometric
quantity generating the evolution. The detailed formulae for $f(y_1,y_2)$
we derive in section \ref{sec:box1}.
\begin{table}[h!]
\begin{center}
\begin{tabular}[c]{|r||c|c|c|} 
\hline 
 & $D_{box}$ & $A_{box}$ & $V_{box}$  \\
 &  [1]    &   [2]         &  [3]      \\
\hline
\hline
\bf Eikonal \rm[0] & 0 & 1 & 2  \\
\hline
\bf Mean Curvature \rm [-1] & -1 & 0 & 1  \\
\hline
\bf Gaussian \rm [-2] & -2 & -1 & 0  \\
\hline
\end{tabular}
\label{table:t1}
\caption{Value of $n$ in equation (\ref{o1}) for 3 component flows in the box equations for 3 observable quantities. In $[]$ brackets we indicated the
power of the maximal size $y_3$ in the given quantities.}
\end{center}
\end{table}

\subsection{Self-similar evolution}
If  we introduce the normalised
observable quantities  
\ben
\bar D(t)  =  D(t)/D(0)\,, \quad
\bar A(t)  =  A(t)/D(0)\,, \quad
\bar V(t)  =  V(t)/D(0),
\een
then we assume that shapes remain self-similar then we have
\ben
\bar A(t)  =  \bar D^2(t)\,, \quad
\bar V(t)  =  \bar D^3(t)
\een
and we have the similar relationship between normalised observable quantities in the box flows
thus our previous results become comparable. We would like to stress that, except for the sphere, shapes
do \emph{not} evolve in a  self-similar manner under these equations; we merely use this assumption
to establish a qualitative correspondence between the results. We summarised the formulae in Table 2
and illustrated them in Figure 2
\begin{table}[ht]
\begin{center}
\begin{tabular}[c]{|r||c|c|c|c|} 
\hline 
 & & & & \\
 & $\bar D(t)$ & $\bar A(t)$ & $\bar V(t)$ & timescale \\
\hline
\hline
\bf Eikonal  & $1-t_e$ & $(1-t_e)^2$ & $(1-t_e)^3$ & $t_e=C_et$  \\
\hline
\bf Mean Curvature & $(1-t_m)^{\frac{1}{2}}$ & $1-t_m$ & $(1-t_m)^{\frac{3}{2}}$ & $t_m=C_mt$  \\
\hline
\bf Gaussian & $(1-t_g)^{\frac{1}{3}}$ & $(1-t_g)^{\frac{2}{3}}$ & $1-t_g$ & $t_g=C_gt$  \\
\hline
\end{tabular}
\label{table:t2}
\caption{Evolution of normalised observable quantities under the assumption that shapes remain self-similar. Equations
apply both in the Bloore PDE and the box flows, only the constants $C_e,C_m,C_g$ differ. In the Bloore flows we have 
$C_e=2\,, C_m=2b(W+4\pi)\,, C_g=4\pi c$. In case of unit spherical particles this yields $C_e=2, C_m=16\pi\approx 50.16, C_g=4\pi\approx 12.56$.
For the same problem in the box equations we get the constants $C_{e,box}=2, C_{m,box}=48, C_{g,box}=12$,
cf. equations (\ref{box_eikonal_size}),(\ref{box_mean_area})and (\ref{box_gauss_volume}), respectively.}
\end{center}
\end{table}

\begin{figure}[ht]
\label{fig:1}
\begin{center}
\includegraphics[width=120 mm]{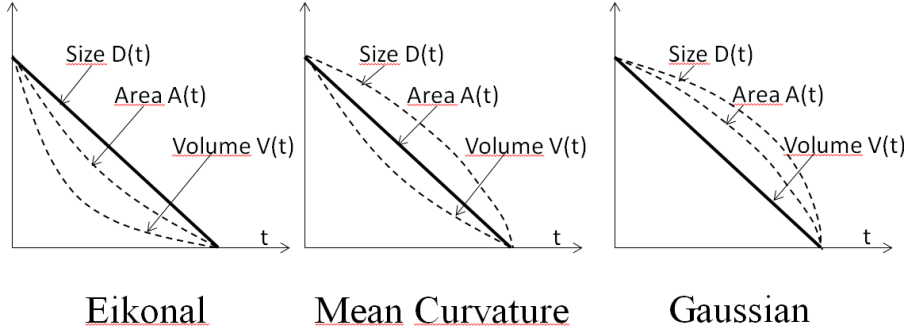}
\end{center}
\caption{Qualitative evolution of observable quantities under the component 
flows.  Each quantity is normalised
by its initial value at $t=0$.  Solid line represent exact result for the PDE,
dashed line represent
qualitative curves under the assumption of self-similar evolution. 
We remark that the box equations
yield the same results under these assumptions. We can observe 
\emph{finite time horizon}
in each case.}\label{fig:components}\end{figure}

\subsection{Observable quantities in the geometrical box flows}\label{sec:box1}

Here we investigate the evolution of observable quantities in the three component flows of (\ref{y_box3D})-(\ref{z_box3D}).

\subsubsection{Eikonal flow}
If both curvature terms are zero then the box flows predict, similarly to the original PDE, linear size diminution, i.e. we have
\ben\label{box_eikonal_size}
\dot D_{box}= 2\dot y_3=-2.
\een
Unlike the original PDE, here we get explicit equations for the area and volume diminution as well:
\bea \label{box_eikonal_area_volume}
\dot A_{box} & = & -16y_3(y_1+y_2+1) \\
\dot V_{box} & = & -8y_3^2(y_1y_2+y_1+y_2)=-A_{box}
\eea
showing that both area and volume diminution is slowing down with time, however, from (\ref{box_eikonal_size}) it is clear
that the particle has a finite time horizon.
If we compare this to the PDE, we can observe that in case of volume evolution the continuous
equations remain valid for polyhedra, however, this is not the case for area evolution.
The naive explanation is that in case of volume evolution the role of the non-smooth
parts (edges, vertices's) is negligible, the bulk of volume
loss is occurring over the smooth (planar) faces and on those
parts the smooth equation is valid. In case of surface area this argument is not true: under the Eikonal action,
polyhedral surface is eliminated at the edges and therefore
the non-smooth effects can not be neglected.

\subsubsection{Mean Curvature Flow}
If the Mean Curvature term dominates the flow then

\bea \label{box_mean_size}
\dot D_{box} & = & 2\dot y_3  =  -\frac{2b}{y_3} \bigl ( \frac{1}{y_1^2} + \frac{1}{y_2^2} \bigr ) \\
\label{box_mean_area}
\dot A_{box} & = & -4b\frac{2y_1^3y_2^3+y_1^3y_2^2+y_1^2y_2^3+y_1^3y_2+y_1y_2^3+y_1y_2^2+y_1^2y_2+2y_1^3+2y_2^3}{y_1^2y_2^2} \\
\dot V_{box} & = & -4by_3\left(\frac{y_1^2+y_2^2+y_1^2y_2^2}{y_1y_2}\right)
\eea
so,  we see that by assuming constant box ratios $y_1,y_2$ linear maximal size is diminishing at an accelerating rate.
Similarly to the PDE, the evolution speed of the surface area is independent of maximal size, however, it is not a constant
but it is approaching a negative constant as the box ratios approach 1. In case of the volume, again assuming constant
box ratios, we see an accelerating decrease.

\subsubsection{Gaussian Flow}
If the Gaussian term dominates the flow then we have

\bea \label{box_gauss_size_area_volume}
\dot D_{box} & = & -\frac{c}{y_3^2}\left(\frac{1}{y_1^2y_2^2}\right) \\
\dot A_{box} & = & - \frac{8c}{y_3}\left(\frac{y_1^4y_2+y_1y_2^4+y_1^4+y_2^4+y_1+y_2}{y_1^2y_2^2}\right)\\
\label{box_gauss_volume}
\dot V_{box} & = & -8c\frac{y_1^3+y_2^3+1}{y_1y_2}
\eea
so we can see that, similar to the PDE, volume evolution is independent of size. Unlike
in the PDE, here the speed is not constant, however, it is approaching a negative constant as
the shape evolves towards the sphere.


\begin{thebibliography}{99}

\bibitem{Bloore} F.~J.~Bloore, The Shape of Pebbles {\it Mathematical Geology} {\bf 9} (1977) 113-122 

\bibitem{DomokosSiposVarkonyi} G.~ Domokos, A.~ Sipo and P.~ V\'arkonyi
Formation of sharp edges and plane areas of asteroids
by polyhedral abrasion {\it Astrophysical Journal} {\bf 699}(2009)  L13-116

\bibitem{Brakke} K.~Brakke, 
{\it The motion of a surface by its mean curvature}
Princeton University Press (1978) 


\bibitem{Sternberg} H.~Sternberg,  Untersuchungen uber Langen-und Querprofil 
geschiebefuhrender Flusse, {\it Z. Bauwes.} {\bf 25} (1875) 486 –506.

\bibitem{DG} G.~Domokos and G.W.~Gibbons, The evolution of pebble shape in space and time
{\it Proceedings of the Royal Society London} \bf 468 \rm (2146) 3059-3079. (2012).

\bibitem{Griffith} A.A. ~Griffith, The phenomena of rupture and flow in solids. 
\it{Philos. T. Roy. Soc. A}, \bf 221, \rm 163-198. (1921).

\bibitem{Kun1} O- ~Tsoungui, D.~Vallet, J-C.~Charmet and S.~Roux, Size effects in single 
grain fragmentation {\it Granular Matter} \bf 2 \rm 19-27. (1999).

\bibitem{VarkonyiDomokos} P.~L.~ V\'arkonyi and G.~Domokos
A general model for collision-based abrasion {\it IMA J. for Applied Math.} \bf 76 \rm 47-56 (2011).


\bibitem{Forsyth} A.~R.~ Forsyth {\it Lectures on the Differential Geometry
of Curves and Surfaces} Cambridge University Press (1920) 



\bibitem{DGS} G.~ Domokos, G.~ W.~ Gibbons, A.~ A.~ Sipos
 Circular, stationary profiles emerging in unidirectional abrasion
[{\tt arXiv:1206.1589 Geophysics (physics.geo-ph)}]


\bibitem{S} A.~A.~ Sipos, G.~Domokos, A.~Wilson and N.~Hovius 
A Discrete Random Model Describing Bedrock Erosion 
{\it Mathematical Geosciences}
{\bf 43}  (2011) 583-591, DOI: 10.1007/s11004-011-9343-8

\bibitem{DSV}G.~ Domokos, A.~ Sipos and P.~ V\'arkonyi
Continuous and discrete models for abrasion processes
{\it Periodica Polytechnica Architecture} {\bf 40} (2009) 3-8
 
\bibitem{Brunt} B. van-Brunt and K. Grant,
Potential applications of Weingarten surfaces in CAGD.
 {\it Computer Aided Geometric Design}
{\bf 13} (1996) 569-582



\bibitem{Firey} W.~.J.~Firey, The shape of worn stones,
{\it Mathematika} {\bf 21}(1974) 1-11

\bibitem{Rayleigh1} Lord Rayleigh, Pebbles, natural and artificial.
Their shape under various conditions of abrasion
{\it Proc R.Soc Lond} {\bf A 181} (1942) 107-118 
 
\bibitem{Rayleigh2} Lord Rayleigh, Pebbles, natural and artificial.
Their shape under various conditions of abrasion
{\it Proc R.Soc Lond} {\bf A 182} (1944) 321-334 

\bibitem{Rayleigh3} Lord Rayleigh, Pebbles of regular shape and
their production in experiment {\it Nature} {\bf 154}
(1944) 161-171

\bibitem{Archard} J.~F.~Archard and W.~Hirst, The wear of metals
 under unlubricated conditions {\it Proc R.Soc Lond} {\bf A 236} (1956) 397-416


\bibitem{Tso} K.~Tso, Deforming a hypersurface
 by its Gauss-Kronecker- curvature, {\it Comm. Pure Appl.
Math} {\bf 38} (1985) 867-882


\bibitem{Andrews} B.~Andrews, Guass curvature flow:the fate of 
rolling stones {\it Invet. Math.} {\bf 138} (1999) 151-161

\bibitem{Andrews1} B.~Andrews, Contraction of convex
hypersurfaces in Euclidean space, {\it Cal Var }{\bf 2} (1994) 151-171


\bibitem{Chow} B. Chow, Deforming convex hypersurfaces
by the n-th root of the Gaussian  curvature, {\it J Diff Geom}
{\bf 22} (1985) 117-138 

\bibitem{Huisken} G.~Huisken, Flow by mean curvature
of convex surfaces into spheres {\it J Diff Geom} {\bf 20}(1984) 27-266



\bibitem{RCD} F.~Rhines, K.~Craig, and R.~ Dehoff,
Mechanism of steady-state grain growth in aluminium,
{\it Metallurgy and Materials Transactions} (1974) 413-425


\bibitem{Schnurer} O.~C.~Schn\"urer, 
Surfaces expanding by the inverse Gauss curvature flow
{\it Journal f\"ur die reine und angewandte 
Mathematik (Crelles Journal)}
{\bf 600} 2006) 117-134 {\tt [arXiv:math/0412297]}


\bibitem{DomokosSiposVarkonyi2}G.~ Domokos, A.~ Sipos and P.~ V\'arkonyi
Continuous and discrete models for abrasion processes
{\it Periodica Polytechnica Architecture} {\bf 40} (2009) 3-8


\bibitem{SchneiderWeil} R. ~Schneider and W.~ Weil, 
{\it Stochastic and Integral Geometry} Springer-Verlag (2008)  

\bibitem{KPZ}
 M.~Kardar, G.~Parisi and Y.\-C.~Zhang {\it Phys Rev Lett }{\bf 56}(1986) 
889-892
\bibitem{Marsilli} M.~Marsilli, A.~Maritan, F. Toigo and J.B.~Banavar,
Stochastic growth equations and reparameterization
invariance {\it Rev Mod Phys} {\bf 68} (1996) 963-983


\bibitem{Maritan} A.~Maritan, F.~Toigo, J.~ Koplik and J.R.~ Banavar,
Dynamics of Growing Interfaces {\it Phys Rev Lett} {\bf 69} (1992) 3193-3195

\bibitem{Batchelor} M.~T.~ Batchelora , R~.V~. Burneb , B~.I~. Henry, 
and S.D. Watt,   Deterministic KPZ model for stromatolite laminae,  {\it Physica A} {\bf 282} (2000)  123-136

\bibitem{Monge} G.~Monge, 
{\it Application de l'anayse \`a la g\'eom\'etrie } 
Paris (1807)  reprinted by ellipses (1994) 
\bibitem{Goldman} R.~Goldman, Curvature formulas for implicit curves and surfaces {\it Computer Aided Geometric Design }{\bf 22}  (2005) 632-658

\bibitem{Williams} Szab\'o, T. Fityus, S. and Domokos, G: Abrasion model of downstream changes in grain shape and size along the Williams River, Australia.
{\it J. Geophysical Research/Earth Surface}, Submitted.

\bibitem{Weibull} W. ~Weibull, A statistical theory of the strength of materials, Roy. {\it Swed. Inst. Eng. Res.} \bf 151 \rm  (1939).

\end{thebibliography}
\end{document}